# Evaluation of Parkinson's disease with early diagnosis using single-channel EEG features and auditory cognitive assessment

Lior Molcho[†1*], Neta B. Maimon[†1,2], Neomi Hezi[5], Talya Zeimer[1], Nathan Intrator[1,3,4], Tanya Gurevich[4,5,6]

[1] Neurosteer Inc, NYC, New York, USA

[2] Department of Musicology, The Hebrew University of Jerusalem, Jerusalem, Israel

[3] Blavatnik School of Computer Science, Tel-Aviv University, Tel-Aviv, Israel

[4] Sagol School of Neuroscience, Tel Aviv University, Tel Aviv, Israel

[5] Movement Disorders Unit, Neurological Institute, Tel Aviv Sourasky Medical Center, Israel

[6] Faculty of Medicine, Tel-Aviv University, Israel

## Abstract

**Background**: Parkinson's disease' (PD) early clinical signs are subtle, making diagnosis challenging. F-DOPA PET imaging offers a reliable measure of dopaminergic function commonly used for early PD diagnosis. Machine-learning (ML) based EEG features may provide a noninvasive, low-cost approach to support PD diagnosis. This study aims to evaluate the ability to predict F-DOPA results and differentiate between PD and non-PD patients using ML-based EEG features extracted from a single-channel EEG during an auditory cognitive assessment.

**Methods**: The study included data collected from participants who underwent an F-DOPA PET scan as part of their standard treatment ($n = 32$) and additional cognitively healthy controls ($n = 20$). Participants performed an auditory cognitive assessment while being recorded with a single-channel EEG by Neurosteer. EEG data processing involved wavelet-packet decomposition and machine learning for feature extraction. Initially, a prediction model was developed to predict one-third of the undisclosed F-DOPA test results. Then, generalized linear mixed models (LMM) were calculated to distinguish between PD and non-PD subjects based on several EEG variables, including frequency bands and ML-based EEG features (A0 and L1) previously associated with cognitive function in separate datasets.

**Results**: The prediction model accurately labeled patients with unrevealed scores as positive F-DOPA. Novel EEG feature A0 and the Delta band showed significant separation between study groups, with healthy controls exhibiting higher activity than PD patients. EEG feature L1 differentiated cognitive load levels in healthy controls, such that resting state L1 activity was significantly lower compared to high-cognitive load conditions. This effect was not observed in the PD group, suggesting the expected difference between high cognitive load and resting state is lacking in PD patients.

**Conclusion**: This study successfully demonstrated the ability to separate patients holding positive vs. negative F-DOPA PET results with an easy-to-use single-channel EEG during an auditory cognitive assessment. Future longitudinal studies should further explore the potential utility of this tool for early PD diagnosis and as a potential biomarker in PD.

# 1    Introduction

Diagnosing Parkinson's Disease (PD) at early stages may be challenging as clinical signs can be subtle, inconclusive, and require differentiation from other disorders. To validate their diagnosis in the early stages, clinicians utilize objective biomarkers of dopaminergic degeneration. Positron emission tomography (PET) scans with [18F]-6-fluoro-L-3,4-dihydroxyphenylalanine (F-DOPA) is an established FDA-approved technique for PD diagnosis (1). However, PET scans have limited availability in clinics due to their large size and high costs. Furthermore, the F-DOPA test is invasive, requires trained personnel, and is difficult to administer. As clinical trials strive to identify disease-modifying treatments for PD, new biomarkers are needed for early-diagnosis validation, with devices at lower costs that can be widely used.

Electroencephalographic (EEG) signals have been extensively studied for over a century and are generally used to investigate cortical and subcortical functionality (2). EEG offers a low-cost and non-invasive means of directly measuring neural activity, which can be analyzed in various dimensions, including time, space, frequency, power, and phase, each reflecting specific neurophysiological mechanisms (3). Advancements in ML and signal processing techniques, such as multi-taper analysis applied to accurately extract various dynamic EEG rhythms (4,5), have significantly contributed to extracting useful information from raw EEG signals (6). Novel techniques can exploit the vast amount of information on time-frequency processes in a single recording (7,8).

Since the loss of dopaminergic neurons affects multiple brain networks, EEG could serve as a research tool in PD (3). Quantitative EEG (qEEG) provides a reliable and widely available measurement that could yield biomarkers for disease severity in PD patients (9). Generally, the incidence of EEG abnormalities in PD patients is higher than in healthy elderly individuals, with the most common alteration being generalized slowing of the EEG (10,11). Some research is available regarding PD diagnosis; for instance, coherence function (CF) has been hypothesized to be a relevant tool for detecting early PD signs (12). CF is related to cortical dynamic imbalances and measures linear dependence through the frequency domain between a pair of electrodes placed on the scalp (13). Coherence can detect changes in functional and effective cortical interconnections that occur in the initial onset of PD (14). Indeed, previous studies have reported that non-linear analysis of EEG signals, particularly machine learning (ML) methods, can extract features that could potentially serve as PD biomarkers (15–21). A recent study published results discriminating early-stage PD from healthy brain function using multi-EEG event-related potentials (ERPs) combined with brain network analytics and ML tools while participants performed an auditory cognitive assessment (22).

In this pilot study, we evaluated the ability of an easy-to-use single-channel EEG system (by Neurosteer®) combined with an auditory cognitive assessment, to detect electrical activity changes caused by PD. Past research indicates that capturing EEG data during active participation in cognitive and auditory tasks can reveal unique features, potentially enhancing the discrimination power of brain states (23). In line with this idea, we utilized an auditory assessment with musical stimuli that has been previously employed to distinguish between cognitive decline and healthy senior participants (24,25). The objective of the present study was to assess the capability of features extracted from a single-channel EEG, conducted with auditory stimulation, to differentiate between positive and negative F-DOPA PET results to be able to potentially discriminate between PD and non-PD populations.

## 2 Materials and Methods

### 2.1 Participants

This study included 32 participants (11 females) with a mean age of 64.15 (SD = 12.3), all holding a valid F-DOPA PET scan results obtained as part of their standard care due to clinical symptoms suspicious of early PD. Additionally, 26 age-matched, cognitively healthy individuals (11 females) with a mean age of 66.19 (SD = 6.49) served as controls. The entire cohort of 52 participants underwent assessment at rest, and then an auditory cognitive assessment, while their brain activity was recorded using a single-channel electroencephalogram (EEG) by Neurosteer. Informed consent was obtained from each participant before their involvement in the study.

### 2.1.1 Participants with F-DOPA PET results

Participants with F-DOPA PET results were recruited from the Movement Disorders Unit at Tel Aviv Sourasky Medical Center if they had an MMSE score higher than 24 and could hear, read, and understand instructions for the Informed Consent Form (ICF) discussion as well as for the auditory assessment tasks. Individuals with compromised scalp or skull integrity, facial or forehead skin irritation, hearing loss, cognitive decline and a history of severe drug abuse were excluded from the study.

Ethical approval for data collection was obtained from the Ethics Committee (EC) of Tel Aviv Sourasky Medical Center (Ichilov) on June 07, 2021. Israeli Ministry of Health (MOH) registry number: MOH_2021-06-02_010019.

### 2.1.2 Healthy Participants

Twenty cognitively healthy age-matched participants were selected from a separate dataset to serve as a control group in the study and balance the group sizes for data analysis purposes.

Ethical approval for the collection of this data was granted by the Ethics Committee (EC) of Dorot Geriatric Medical Center on September 07, 2020. NIH registry number: NCT04683835.

### 2.1.3 Study Groups

The 52 study participants were initially divided into four groups based on their F-DOPA results (see Figure 1): participants with a positive F-DOPA result ($n = 14$); participants with a negative F-DOPA result ($n = 6$); participants whose label was initially unrevealed in the 'unknown' group ($n = 12$); and healthy age-matched controls (n = 20). These groups were used in building and testing the prediction model.

For the second part of the analysis, the labels were revealed, and participants were added to the relevant groups: the healthy age-matched controls were combined with the negative group to form the 'healthy' group ($n = 26$), which was compared to the 'PD patients' group consisting of patients with a positive F-DOPA result ($n = 26$).

To ensure that the groups were well-balanced in terms of age, gender, and MMSE scores, we compared the mean ages of each group in total and separately for males and females. Additionally, we compared the age and MMSE scores of each group between males and females. These comparisons were conducted using the Welch Two Sample t-test.

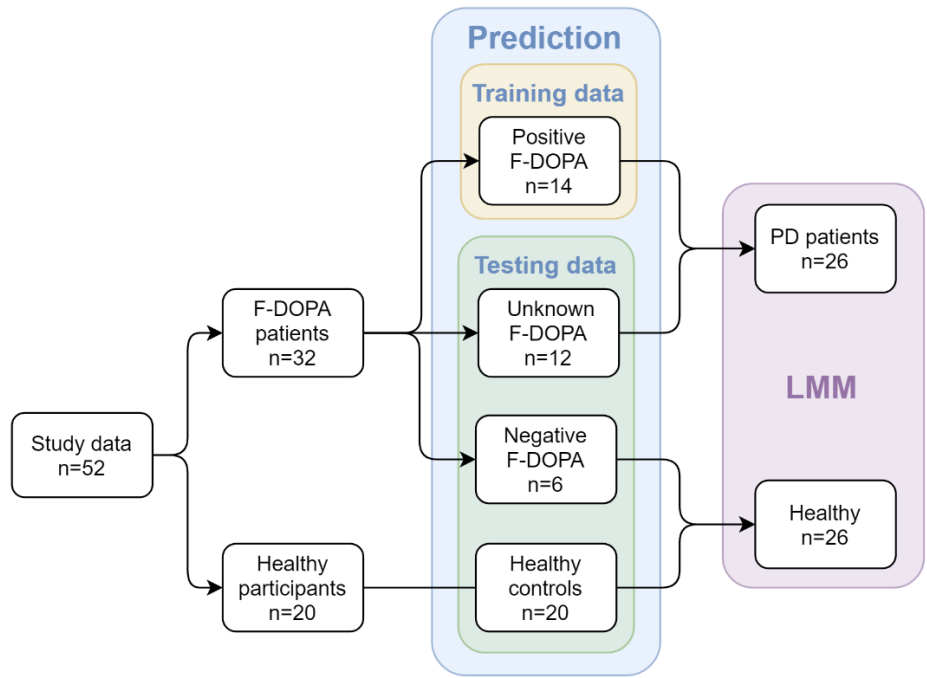

**Figure 1.** Study design and groups at each analysis stage. The study included patients with valid F-DOPA results and age-matched healthy participants as controls. In the first part of the analysis, a prediction model was used; patients with an initial positive F-DOPA score were included in the model's training data, while patients with an initial negative F-DOPA score, patients with unrevealed test scores, and healthy controls made up the testing data. In the second part of the analysis (after the results were revealed), a Linear Mixed Model (LMM) was utilized to compare between two combined groups: the PD-patients group ($n = 26$, patients with positive F-DOPA results) and the healthy group ($n = 26$, patients with negative F-DOPA results along with healthy controls).

## 2.2 EEG device

EEG recordings were conducted using the Neurosteer® EEG Recorder. A medical-grade patch with three electrodes was placed on the participant's forehead, using dry gel for optimal signal transduction. The non-invasive monopolar electrodes were positioned at the prefrontal regions, generating a single EEG channel from the difference between Fp1 and Fp2 in the international 10/20 electrode system. A reference electrode was placed at Fpz, with an input range of ±25 mV (input noise < 30nVrms). EEG-electrode contact impedances were kept below 12 kΩ, as measured by a portable impedance meter (EZM4A, Grass Instrument Co., USA). Data was digitized in continuous recording mode at a 500-Hz sampling frequency.

A trained research assistant monitored each participant during recordings to minimize muscle artifacts. Participants were instructed to avoid facial muscle movements during recordings, and the operator alerted them if they exhibited increased muscle or ocular movements. It's worth noting that the differential input and high common-mode rejection ratio (CMRR) help to remove motion artifacts and line noise (26).

## 2.3  Procedure

### 2.3.1 EEG recording and auditory assessment protocol

The recording room was quiet and well-lit. The research assistant prepared the sanitized Neurosteer EEG recorder equipment (electrode patch, sensor, EEG monitor, clicker) for use. The electrode was placed on the participant's forehead, and the recording began. Each participant sat during the assessment and received instructions through a loudspeaker connected to the EEG monitor. The entire recording session typically lasted 20-30 minutes. The cognitive assessment battery was pre-recorded and included tasks such as musical detection, musical n-back, and resting state tasks. The research assistant provided general instructions to participants before starting the tasks, and further explanations were kept to a minimum to avoid bias. A few minutes of baseline activity were recorded for each participant to ensure accurate testing.

The auditory cognitive assessment took approximately 15 minutes to complete and included a simple auditory detection task with two difficulty levels (low and high), an auditory musical n-back task, and a resting-state task.

#### 2.3.1.1 Cognitive Tasks

In this study, we employed a previously described auditory detection task (24), an auditory n-back task, and resting state tasks.

In the detection task, participants listened to a sequence of melodies played by a violin, a trumpet, and a flute. They were given a clicker to respond to the stimuli. At the beginning of each block, auditory instructions specified an instrument for which the participant would click once. The click response was only for "yes" trials when the indicated instrument's melody played. The task included two difficulty levels to examine increasing cognitive load. Detection level 1 featured the same melody played for three seconds and repeated throughout the entire block. Participants were asked to click once as quickly as possible for each melody repetition. This level consisted of three 90-second trials (one for each instrument), with 5-6 instances of each melody and 10-18 seconds of silence in between. Detection level 2 presented the same melodies played for 1.5 seconds, with all three instruments appearing in the same block. Participants were asked to click only for a specific instrument within the block and to ignore the other melodies. Each trial included 6-8 melodies, with 8-14 seconds of silence in between and 2-3 instances of the target stimulus.

In the n-back task, participants were presented with a sequence of melodies played by different instruments, and they used the same clicker to respond to the stimuli. This task also included two difficulty levels (0-back and 1-back) to examine increasing cognitive load. A set of melodies (played by a violin, a trumpet, and a flute) was played in a different order for each block, and participants were asked to click a button when the melody repeated n-trials ago (based on the block level). In the 0-back level, participants clicked the button each time a melody was heard. This level included one 90-second block with 9 trials (instances of melody playing), each melody played for 1.5 seconds and 6-11 seconds of silence in between. In the 1-back level, participants clicked the button each time a melody repeated itself ($n = 1$). This level included two 90-second blocks with 12-14 trials (instances of melody playing), each melody played for 1.5 seconds and 4-6 seconds of silence in between. In each block, 30%-40% of the trials were the target stimulus, where the melody repeated itself, and the participant was expected to click the button. The resting state tasks consisted of two blocks: one with 45 seconds and the other with 60 seconds of resting state recording.

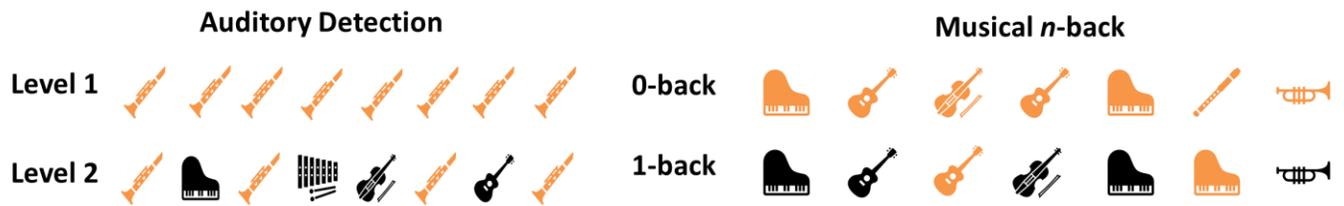

**Figure 2**. A visual representation of the two cognitive tasks used in this study is provided. Auditory Detection (left): Level 1 features the same melody played by the same musical instrument several times, and the participant is asked to click each time the melody is played. Level 2 presents melodies played by different instruments, and the participant is asked to click only when a melody by a specific instrument is played (in this example, the flute melody). Musical n-back (right): Levels 1 and 2 showcase melodies played by various instruments. In Level 1, the participant is asked to click whenever a melody is played, while in Level 2, the participant is asked to click only when a melody immediately repeats itself (regardless of which melody is played).

### 2.3.2 Signal processing

The EEG signal was decomposed into multiple components using harmonic analysis mathematical models (5,27), and ML methods were employed on the components to extract higher-level EEG features. The full technical specifications for signal processing can be found in Molcho et al., 2022 (24). In summary, the Neurosteer® signal-processing algorithm analyzes EEG data using a time/frequency wavelet-packet analysis. This analysis, previously conducted on a separate dataset of EEG recordings, identified an optimal orthogonal basis decomposition from a large collection of wavelet packet atoms, optimized for that set of recordings using the Best Basis algorithm (28). This basis results in a new representation of 121 optimized components called Brain Activity Features (BAFs). Each BAF consists of time-varying fundamental frequencies and their harmonics.

The BAFs are calculated over a 4-second window, which contains 2,048 time elements due to the 500 Hz sampling frequency. In this window, each BAF is a convolution of a time/frequency wavelet packet atom, allowing for a signal that can vary in frequency over the 4-second window, such as a chirp. The window is then advanced by one second, similar to a moving window spectrogram with 75% overlap, and the calculation is repeated for the new 4-second window. The EEG power spectrum is obtained using a fast Fourier transform (FFT) of the EEG signals within a 4-second window.

## 3   Results

### 3.1   Demographic results

To ensure that the groups were well-balanced we compared some demographic characteristics of each group. First, for the positive vs. negative patients within the F-DOPA group, we compared the F-DOPA test age, symptoms onset age, and difference between the F-DOPA test and the auditory assessment task (in years), see Table 1 for details. No significant differences were found between these sub-groups (all $p$s > 0.05). Due to the small sample size of the negative F-DOPA group ($n$=6), this analysis did not include a division between males and females. For descriptive information about the motor symptoms side see Supplementary Material B, table 1.

| Study Group | Gender | n | Auditory task age, years | F-DOPA test age, years | Symptoms onset age, years | F-DOPA – Neurosteer difference, years |
|---|---|---|---|---|---|---|
| F-DOPA Negative | Female | 4 | 61.77 (7.97) | 62.03 (8.30) | 57.79 (9.94) | 0.49 (0.50) |
| | Male | 2 | 74.50 (0.50) | 75.01 (1.00) | 74.50 (0.5) | 0.5 (0.50) |
| F-DOPA Positive | Female | 11 | 65.10 (13.10) | 63.64 (12.85) | 56.41 (12.72) | 1.91 (1.73) |
| | Male | 15 | 63.54 (11.22) | 62.73 (11.41) | 58.79 (11.59) | 1.14 (1.10) |

**Table 1**. F-DOPA clinical information for the groups included in the analysis. Averages are shown for the total number of participants.

All recruited patients completed the auditory assessment tasks, and their EEG data was used. The average age of patients with F-DOPA results was 64.50 (11.73) years, with 47% females and 53% males. The average MMSE score was 29.46 (0.76). The average age of the healthy control group was 66.25 (5.55) years, with 35% females and 65% males. The average MMSE score was 29.15 (0.81). Overall, the mean age was 65.17 (9.79) years, with 42% females and 58% males. No significant differences in age or gender were found between the groups (all $p$s > 0.05). See Table 2 for complete demographic details and results.

| | Groups | PD patients | Healthy |
|---|---|---|---|
| Total | n | 26 | 26 |
| | MMSE | 29.61 (0.57) | 29.07 (0.89) |
| | Age | 64.15 (12.30) | 66.19 (6.49) |
| | Age t-tests | PD patients vs. healthy: $t = 0.54$, $p = 0.58$ | |
| Male | n | 15 | 15 |
| | MMSE | 29.66 (0.61) | 29.06 (0.88) |
| | Age | 63.46 (11.61) | 67.13 (5.89) |
| | Age t-tests | PD patients vs. healthy (male): $t = 1.09$, $p = 0.28$ | |
| Female | n | 11 | 11 |
| | MMSE | 29.54 (0.52) | 29.09 (0.94) |
| | Age | 65.09 (13.71) | 64.90 (7.34) |
| | Age t-tests | PD patients vs. healthy (female): $t = -0.03$, $p = 0.96$ | |
| Age males vs. females | | $t = -0.32$, $p = 0.74$ | $t = 0.85$, $p = 0.39$ |

**Table 2**. Demographic information for the groups included in the analysis is presented. Averages are shown for the total number of participants, as well as for males and females separately. $t$ and $p$ values of the comparisons between mean ages of the study groups are displayed for the total, and for males and females separately. Additionally, $t$ and $p$ values of the comparisons between age and MMSE scores for each gender are provided in the last rows.

## 3.2 Prediction model of F-DOPA results

For a full description of the prediction model methodology see Supplementary Materials A. In the initial phase of data analysis, our primary objective was to develop a predictor capable of accurately classifying and predicting F-DOPA test results. The prediction model was formulated using machine learning (ML) methods applied to the extracted BAFs. As an integral component of the study design, one-third of the F-DOPA results were intentionally undisclosed to evaluate the prediction model. The process of developing such a predictor entails three steps: (1) identifying the feature representation from which the prediction is derived, (2) determining the type of data to be utilized in training the predictor, and (3) ascertaining the model family from which a predictor will be selected. In this pilot study, our primary focus was on identifying the type of representation that could yield a meaningful prediction. Consequently, we maintained the other two factors as constants, as detailed below.

To that end, we initially tested whether a feature representation based on connectivity between the BAFs was useful. Connectivity and causality Analysis has been used successfully in the context of neuroscience (29). In this study, we adopted the approach used by Friston et al., 2003 (29), applying it to the components extracted from single-channel EEG data rather than multiple electrodes or multiple fMRI regions (see full details in the Appendix). We employed connectivity-based representation and performed dimensionality reduction using principal components analysis (PCA) (30).

The PCA-derived reduced-dimensionality representation was used for training and testing the prediction model. We used a previously collected dataset, which included data from healthy participants and patients with PD performing similar auditory tasks, in conjunction with the positive F-DOPA labels collected in this study to serve as training data for the predictor. The testing data comprised of participants with undisclosed F-DOPA results, participants with negative F-DOPA, and healthy controls.

The prediction model was based on an ensemble of ridge regression (31). Ridge regression extends linear regression by modifying the loss function to minimize the model's complexity, introducing a constraint on the coefficients through a penalty factor equivalent to the square of the magnitude of the coefficients. The ensemble predictor consisted of 10 logistic regression predictors with regularization terms ranging from 1 to 10 (32). Studies have shown that ensembles with strong regularization values can mitigate noise in the data and produce better predictors (33).

The trained PCA model with ridge regression yielded a score between -1 and 1 for each participant, corresponding to a predicted test result label. A separating cutoff score of 0 was set, with data points higher than 0 classified as positive F-DOPA and those lower than 0 classified as negative F-DOPA. The prediction model labels were compared to the actual test labels to determine the model's accuracy in classifying the 12 unknown patients and accurately classifying other groups as either negative or positive.

Due to a tendency for positive bias among patients referred for F-DOPA scans as part of standard care, the majority of collected F-DOPA results were positive. To include negative results, the six patients with negative F-DOPA results were also considered as part of the testing data. Since all 12 patients in the unknown group were eventually classified as positive, we performed additional quantitative analysis using Bayesian Mann-Whitney U Tests to determine the similarity between labeled groups. This follow-up analysis was conducted using a data augmentation algorithm with 5 chains of 1000 iterations. We report the $BF_{01}$ (i.e., the null hypothesis that H0 is not different from H1) of the Bayesian U tests between controls vs. negative and positive vs. unknown, and the $BF_{10}$ (i.e., the hypothesis that H0 is different from H1) of the Bayesian U tests between control vs. positive and control vs. unknown groups. This analysis was performed using JASP 0.11.1.0 software (34) (JASP, Version 0.17).

### 3.2.1 Prediction model results

Figure 3 depicts the prediction model results. All 12 patients in the 'Unknown' group were classified as having a positive F-DOPA result based on the prediction model (i.e., all predictor results were >0). Moreover, the predictor assigned negative values to the six patients initially labeled as negative F-DOPA and positive values to all 14 patients initially labeled as positive F-DOPA. The majority of the control group samples received negative values as expected, except for 4 samples in the control group (20%), who received a positive F-DOPA label. It would be of interest to follow these four individuals and test if there was a pre-symptomatic detection of dopamine depletion in these subjects.

Bayesian Mann-Whitney U tests revealed strong evidence that the predicted results of the control group differ from positive F-DOPA patients ($BF_{10}$ = 121.88, W = 385, $R^2$ = 1.04), and presented moderate evidence of similarity to the negatively labeled F-DOPA group ($BF_{01}$ = 2.97, W = 116, $R^2$ = 1.21). The group with unknown labels, who were all given positive predictor results, was strongly evident to differ from the control group ($BF_{10}$ = 149.48, W = 550, $R^2$ = 1.032), and showed moderate evidence of similarity to the positive group ($BF_{01}$ = 2.145, W = 98, $R^2$ = 1). For all U tests outputs and figures, see Supplementary Materials B.

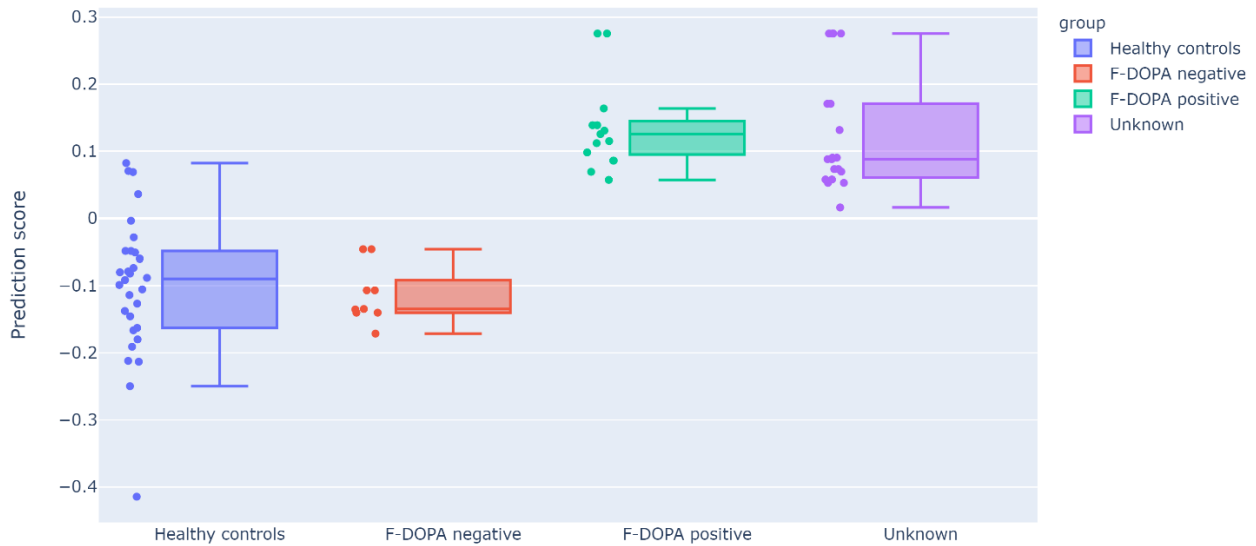

**Figure 3.** Results of the prediction model. The prediction scores (y-axis) cutoff between positive and negative labels is 0 (data points with prediction scores higher than 0 are classified as positive F-DOPA, whereas prediction scores lower than 0 are classified as negative F-DOPA). The study data are displayed in the graph as individual sample points and as F-DOPA groups: positive F-DOPA results, negative F-DOPA results, healthy age-matched controls, and initially unrevealed unknown F-DOPA results.

### 3.3 High-level features and Mixed Linear Models (LMM) analysis

In the second phase of data analysis, our focus was on evaluating the ability of previously extracted high-level EEG features and conventional frequency bands to differentiate between PD patients and healthy controls based on the auditory assessment protocol. To achieve this, we employed mixed linear models to evaluate the associations between EEG variables, groups, and cognitive load levels.

### 3.3.1 High-level EEG features

The EEG features were previously generated using machine learning (ML) techniques applied to the Brain Activity Features (BAFs) from labeled datasets previously collected by Neurosteer. Specifically, EEG features A0 and L1, employed in this study, were calculated using the linear discriminant analysis (LDA) technique (35). The LDA technique aims to identify an optimal linear transformation that maximizes class separability.

Data analysis included the activity of EEG features A0 and L1, normalized to a scale of 0-100. The EEG variables were calculated every second from a moving window of four seconds, and the mean activity per condition was incorporated into the analyses.

### 3.3.2 Frequency bands

The electrophysiological dependent variables incorporated the power spectral density. Absolute power values were converted to logarithm base 10, resulting in values expressed in dBµV. Among the frequency bands, Delta (0.5-4 Hz) and Theta (4-7 Hz) were included. Preliminary tests indicated that the other frequency bands, such as Alpha (8-15 Hz), Beta (16-31 Hz), and lower Gamma (32-45 Hz), did not demonstrate any significant correlations or differences in the current data.

### 3.3.3 LMM analysis comparing PD patients and Healthy Controls

In order to detect differences between PD patients and healthy controls, we employed a general linear mixed model (GLMM) (36), which incorporates both fixed and random effects. This model was preferred over the simpler GLM due to the relatively small sample size, as the GLMM accounts for the random slope for each participant. The model included the fixed within-participant variable of cognitive load level, as well as the group as a between-participants variable.

The group variable consisted of two levels: 'PD Patients' (patients with positive F-DOPA results) and 'Healthy Controls' (comprising both patients with negative F-DOPA results and healthy age-matched controls). As an initial validation, student t-tests were performed on each EEG variable between the subjects in the 'Healthy Controls' group to ensure there were no inherent differences in EEG activity between the two sub-groups (i.e., patients with negative F-DOPA results vs. healthy age-matched controls).

The cognitive load variable was an ordinal variable, coded linearly according to the task cognitive load level (from low to high) as follows: resting state = 0; detection level 1 = 1; 0-back = 1; detection level 2 = 2; and 1-back = 2. The model included the samples per participant per task (i.e., samples per second of activation) as a random slope. For models that demonstrated a significant main effect of cognitive load, post-hoc analyses were conducted, comparing possible pairwise combinations of cognitive load levels for each group (i.e., healthy vs. PD), using the Benjamini-Hochberg correction (37) for multiple comparisons. The significance level for all analyses was set to $p<0.05$. All analyses were conducted using RStudio version 1.4.1717 (38).

### 3.3.4 LMM results – comparing EEG variables between PD patient and healthy controls

### 3.3.4.1 Initial Validation

To rule out any intrinsic differences within the 'healthy controls' group, we compared between the subgroups composing the group: patients with negative F-DOPA results and the healthy age-matched patients. No significant differences were found for any of the EEG variables ($p = 0.249$, $p = 0.64$, $p = 0.3$ and $p = 0.406$ for Delta, Theta, A0 and L1, respectively).

### 3.3.4.2 LMM Models

For a full description of models' outputs see Table 3. Delta and A0 showed higher mean activity for the healthy controls compared to the PD patients ($p = 0.01$ and $p = 0.003$, respectively). Cognitive load ordinal effect reached significance for L1 ($p = 0.043$). Paired t-test analysis revealed that for the healthy group, L1 activity during the resting state task was significantly lower than during the high-cognitive load condition (adjusted $p = 0.022$), whereas in the PD group, no significant difference was found in L1 activity between any of the cognitive load conditions (see Table 4 and Figure 4).

|       | Fixed Effect   | Coef.  | Std.Err. | z      | P>\|z\|    | [0.025 | 0.975] |
|-------|----------------|--------|----------|--------|----------|--------|--------|
| Delta | Intercept      | 2.24   | 0.87     | 2.58   | 0.01     | 0.54   | 3.94   |
|       | Group          | 2.99   | 1.16     | 2.58   | **0.01** | 0.72   | 5.25   |
|       | Cognitive Load | 0.12   | 0.26     | 0.45   | 0.651    | -0.39  | 0.62   |
| Theta | Intercept      | -7.38  | 0.69     | -10.70 | <0.001   | -8.73  | -6.03  |
|       | Group          | 1.24   | 0.96     | 1.28   | 0.199    | -0.65  | 3.12   |
|       | Cognitive Load | 0.21   | 0.17     | 1.21   | 0.226    | -0.13  | 0.54   |
| A0    | Intercept      | 72.72  | 1.59     | 45.74  | <0.001   | 69.60  | 75.84  |
|       | Group          | 6.30   | 2.13     | 2.97   | **0.003**| 2.14   | 10.47  |
|       | Cognitive Load | 0.24   | 0.21     | 1.13   | 0.258    | -0.18  | 0.66   |
| L1    | Intercept      | 48.17  | 1.51     | 31.84  | <0.001   | 45.20  | 51.13  |
|       | Group          | -0.30  | 2.15     | -0.14  | 0.889    | -4.51  | 3.91   |
|       | Cognitive Load | 0.65   | 0.32     | 2.02   | **0.043**| 0.02   | 1.28   |

**Table 3**. Fixed effect coefficients, standard error, *z*-values, *p*-values, and 95% confidence interval outputs from the LMMs conducted on EEG features, with group (healthy controls vs. Parkinson's patients), and cognitive load (resting state vs. low-load vs. high-load) coded as numeric variable.

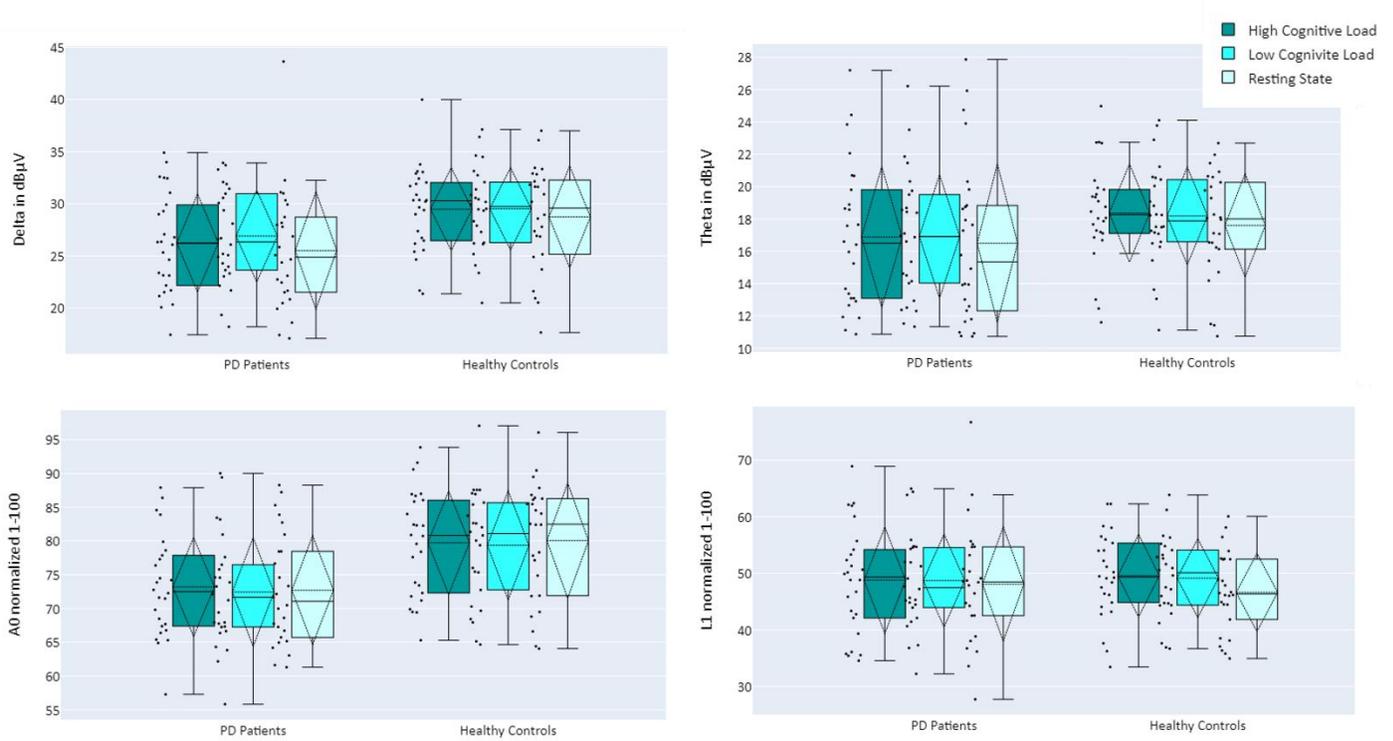

**Figure 4**. Mean activity of EEG features Delta (top left), Theta (top right), A0 (bottom left), and L1 (bottom right), comparing PD Patients (left) and Healthy Controls (right) during performance of cognitive tasks, as a function of cognitive load: high cognitive load (dark turquoise), low cognitive load (turquoise), and resting state (light turquoise).

| Group | comparison | t value | p value | p adj BH |
|---|---|---|---|---|
| Healthy Controls L1 activity | (high, low) | 2.92 | 0.007 | **0.022** |
| | (mid, low) | 2.11 | 0.045 | 0.068 |
| | (mid, high) | -0.61 | 0.544 | 0.544 |
| PD Patients L1 activity | (high, low) | 0.60 | 0.549 | 1.649 |
| | (mid, low) | 0.52 | 0.607 | 0.911 |
| | (mid, high) | -0.08 | 0.933 | 0.933 |

**Table 4**. *t* values, *p* values and *p* BH adjusted values of the pairwise comparisons of the L1 activity per each group, between the three cognitive load conditions: high-load, low-load and resting state.

# 4 Discussion

F-DOPA PET is recognized as a highly sensitive diagnostic tool for parkinsonism (39). Reliability studies employing bilateral Intraclass Correlation Coefficients (ICCs) in test-retest designs have demonstrated that F-DOPA imaging offers a reliable measure of dopaminergic function in the striatum (40). However, F-DOPA PET scans are both costly and invasive. The identification of valid, readily available biomarkers for early diagnosis and disease progression in PD remains a significant unmet need in PD research (41). It has been proposed that artificial neural networks and machine learning (ML) models may contribute to the discovery of specific prognostic biomarkers. Furthermore, functional connectivity and network analyses may hold potential as novel specific biomarkers (9). The availability of such objective biomarkers for disease severity and progression in PD could directly facilitate early diagnosis of nonmotor symptoms, provide a more reliable prognosis, and enable objective monitoring of progression, both in the context of clinical practice and clinical trials.

Following this approach, two novel EEG features extracted using ML methods were used in this study. Notably, the EEG features, A0 and L1, were derived from different datasets than the data analyzed in the present study. Therefore, the same weight matrices previously identified were utilized to transform the data obtained in this study, significantly reducing the risk of model overfitting. Recently published research presented results regarding the two EEG features. A study conducted on young healthy participants demonstrated that activity of EEG feature L1 increased with rising cognitive load levels, as manipulated by a numeric n-back task (42). Furthermore, L1 activity decreased during the performance of an arithmetic task with external visual interruptions (43). In the clinical population, activity of EEG feature A0 correlated with cognitive load levels in healthy young participants and exhibited a correlation to MMSE scores of senior participants in different cognitive states (24). Moreover, A0 demonstrated the ability to differentiate between cognitively impaired and healthy participants, with higher sensitivity to advanced cognitive decline states and the capacity to detect subtle changes in brain activity functioning, even when impairment is mild. The single-channel EEG system used to collect the data, combined with the auditory assessment protocol, have been used in previous studies involving elderly populations (24,25) and was reported to be well-tolerated among senior participants. In this pilot study, a prediction model based on these extracted features was evaluated for its ability to accurately label initially unrevealed results of one-third of the F-DOPA scans. We demonstrated the capability of these ML-based features extracted from the single-channel EEG to differentiate between patients with positive vs. negative F-DOPA PET results and distinguish between PD and non-PD populations.

Specifically, the predictor successfully identified that the 12 unrevealed test results should be labeled as positive. Additionally, the predictor accurately labeled the results of the negative as well as the positive F-DOPA scans initially known. Most of the control group received a negative label as expected, except for 4 samples in the control group (20%), who received positive F-DOPA labels. This may be within the margin of error, or these participants may be prodromal PD patients. A follow-up longitudinal study following these patients would be beneficial to corroborate this. Nevertheless, the preliminary prediction results presented here support the notion of the power of prediction for individual patients rather than analysis of a group of patients. The supplementary Bayesian analysis further substantiated the predictor's accuracy, providing evidence that the predictor output for F-DOPA-positive patients significantly diverged from that of healthy controls and the patients with negative results. In contrast, the output for F-DOPA-positive patients was indistinguishable from that of the 12 previously unknown patients, all subsequently confirmed as F-DOPA-positive. As the predictor was specifically engineered to distinguish between F-DOPA-positive and F-DOPA-negative outcomes, no notable differences were observed between the nine F-DOPA-negative patients and the control group participants. Nonetheless, the possibility of existing disparities in brain activity between the F-DOPA-negative patients who were referred to preform an F-DOPA test due to some clinical manifestations and the healthy controls cannot be dismissed, and the brain activity of the symptomatic F-DOPA negative group should be further investigated. This notion is supported by the raw data underlying the predictor's development, as depicted in Supplementary Materials A, Figure 2.

Upon revealing all F-DOPA results, further analysis was conducted to compare the electrophysiological activity of PD patients vs. healthy controls, as well as cognitive load levels (manipulated by different auditory cognitive tasks). Results indicate that the Delta band and EEG feature A0 differentiate between the groups: Delta and A0 exhibited lower activity for the PD patients compared to healthy controls. This difference was more pronounced for A0 than Delta, suggesting that A0 may be more sensitive to functional changes, a notion supported by the highest separation demonstrated between groups with different levels of cognitive decline in a previous study (24). Finally, the L1 biomarker, which was previously shown to correlate with cognitive load (42), exhibited lower activity in resting state for healthy controls. In contrast, PD patients did not display such decrease in L1 activity in the resting state condition. This finding aligns with previous research regarding resting state activity within PD patients. PD is characterized by higher resting EEG total power compared to healthy controls and slower oscillations in brain activity during resting state – a phenomenon independent of the disease's stage, duration, and severity, and is also resistant to treatment with dopamine (44,45). In conjunction, activation patterns of the two biomarkers – a decrease in A0 mean activity and no difference in L1 between high cognitive load and rest may serve as early indications of PD.

Despite the promising initial results, this study has several limitations. The generalization of the results is restricted due to the small sample size, and further studies with larger cohorts of patients are necessary to validate these preliminary findings. Investigations comparing various other indications for PD, including cerebrospinal fluid (CSF) and blood biomarkers, would also be beneficial in validating the EEG biomarkers and their predictive power. Moreover, the absence of detailed clinical information precludes the performance of in-depth EEG-clinical correlations; We do not have enough data to design a clinical profile of the negative F-DOPA group. Follow-up studies testing the symptomatic patients who received a negative F-DOPA result, as well as the four healthy patients, would greatly contribute to a better understanding of the results. Our approach employs wavelet-packet analysis as a pre-processing step for ML, creating components composed of time-varying fundamental frequencies and their harmonics. These complex time/frequency components of dynamic nature are instrumental in the interpretation of the EEG signal. Future research should explore the utility of this approach in the assessment of neurological disorders. Additionally, examining the potential usefulness of the EEG features presented here in controlled studies characterizing EEG psychogeography in seniors may contribute to understanding the association of these features with basic brain function. This warrants further investigation to evaluate the single-channel EEG with ML analysis as a potential new biomarker in the context of PD.

The fact that a single-channel EEG with auditory stimulation was able to differentiate between patients with positive vs. negative F-DOPA PET results may support the hypothesis that a single-channel EEG could reflect the dopaminergic function of the brain. Furthermore, while F-DOPA PET is based on metabolic function and predominantly reflects dopaminergic deficit, EEG data may potentially represent functional disability due to dopaminergic deficit. Discrimination based on features extracted from a single EEG channel could potentially lead to an objective physiological assessment to aid in the early detection and diagnosis of PD.

# *Supplementary Material A – Appendix I*

## 1    Prediction model

### 1.1    Background

In the context of neuroscience, connectivity analysis was proposed to study the correlation or causality of BOLD activity between different brain regions (1). It was later extended to measure the correlation and causality of EEG brain activity between different electrodes (2), and further extended to perform the connectivity between electrodes after undergoing an independent components analysis (3), source localization analysis (4), or other multiple electrodes transformation such as spectral PCA (5). Thus, connectivity analysis so far, has been between different fMRI regions, or different EEG electrodes, and in general, between multiple sources. Here we present connectivity analysis between components obtained from a single source -- the same EEG channel.

In Parkinson's disease (PD), connectivity analysis proved to be a useful tool to establish underlying pathophysiology and network connectivity related to the motor and non-motor symptoms of the disease. Multiple studies using fMRI-based connectivity analysis give insight into disrupted connectivity in the PD patient population. An abnormal activation of different areas of the motor and resting state networks in patients with early and late-stage PD was shown to be related to cardinal clinical features. For example, it was shown that Parkinson's tremor-related activity first arises in the basal ganglia and is then propagated to the cerebello-thalamo-cortical circuit, determining the role of those circuits in initiating and maintaining tremor symptoms (6). Another study suggests that some of the factors related to PD patients having difficulty achieving automatic movement are less efficient neural coding of movement and failure to shift execution of automatic movements more subcortically, and these changes of effective connectivity become more abnormal as the disorder progresses (7). Additionally, a great body of evidence exists in the literature discussing the decreased functionality in default mode network (DMN) as part of PD progression. For example, resting state fMRI connectivity revealed a disrupted functional integration in cortico-striatal loops in the sensorimotor network in patients with PD (8). Decreased functional connectivity in mesolimbic-striatal and cortico-striatal loops was found in drug-naïve PD patients compared to healthy controls (9). It has also been shown that PD patients with cognitive impairment predominantly showed a reduced connectivity in specific brain regions that are part of the default mode network (10).

EEG connectivity studies have also demonstrated functional connectivity disruptions in PD patients. A recent review exploring over 85 such studies concluded that the main observations were a general slowing of background activity, excessive synchronization of beta activity, and disturbed movement-related gamma oscillations in the Basal-Ganglia and in the cortico-subcortical and cortico-cortical motor loops (11).

### 1.2    Methods

In the prediction model used in this study, we present connectivity analysis that is based on a single EEG source. This is done by first decomposing the signal into multiple components via a time-frequency optimal orthogonal decomposition, then, performing connectivity analysis on the components. We performed an orthogonal decomposition using the Best Basis Algorithm

(12) as described in Molcho et al., 2022 (13). This contrasts with a connectivity analysis based on multiple fMRI regions, or multiple EEG electrodes, as described in the previous section.

We follow the methodology of BOLD connectivity analysis as described in Fristen et al., 1996 (14), but use it on the components decomposed from the single-channel EEG. These components were extracted from earlier collected data of multiple subjects performing multiple tasks. The observations were in the form of 4-second time windows with 3 second overlap. Projecting the 121 components previously extracted on the data collected in this study lead to 121 time series that are updated once per second. From these time series, we obtain a 121x121 connectivity matrix. The matrix can be symmetric and indicate correlations between the different time series or non-symmetric and indicate directed causality between the components. Here we focus on the symmetric correlation matrix.

The matrix of correlations that is obtained from each individual patient can be used as an input to an AI-based predictor that is trained by labeled data previously collected (Figure 3 in manuscript). It can also be used for a group analysis and visualization of the connectivity of each group of patients (Figure 1 Suppl.). To obtain patient population connectivity, we average the correlation matrix over the group members with a specific clinical diagnosis. Given an averaged matrix of correlations between different components, a classical connectivity representation (15), is not possible due to the large number of connections to present: 121x60. We resolve this by converting the correlation matrix into a matrix of distances between components, such that higher correlation reflects a smaller distance, by using:

$$d_{ij} = e^{-\text{corr}(c_i, c_j)}$$

Then, we project the high dimension space of all correlations (121x120/2 dimensions) onto a two-dimensional space while preserving the local distance between every two components as much as possible. This is done via multi-dimensional scaling (MDS) as suggested for fMRI by Friston et al., 1996 (14). Friston calculated the correlation between different regions and then projected via MDS to obtain a distance map that is implied by the correlations. The projection via MDS, provides a practical visualization for the connectivity of a large number of nodes, which in our case refer to the 121 components extracted from the single-channel-EEG.

### 1.3 Results

Figure 1 presents the connectivity maps based on the MDS representation described above. The figure contains the 121 components included in the analysis (extracted from the single-channel-EEG). They are marked by numbers on the graphs and corresponding heat-map colors. The distances between each component to another, represent the correlation between them as obtained by the model. A lower absolute distance between two components in the graph represents higher connectivity between those two components. The figure presents 3 different groups of patients: positive F-DOPA (left, n=24), Negative F-DOPA (middle, n=6), and healthy age-matched controls (right, n=24). A clear component connectivity difference between the F-DOPA positive group and the healthy age-matched group can be seen. Interestingly, the F-DOPA negative patients' connectivity map appears closer to the healthy group in correlation between the components, but with distinct changes which should be further explored.

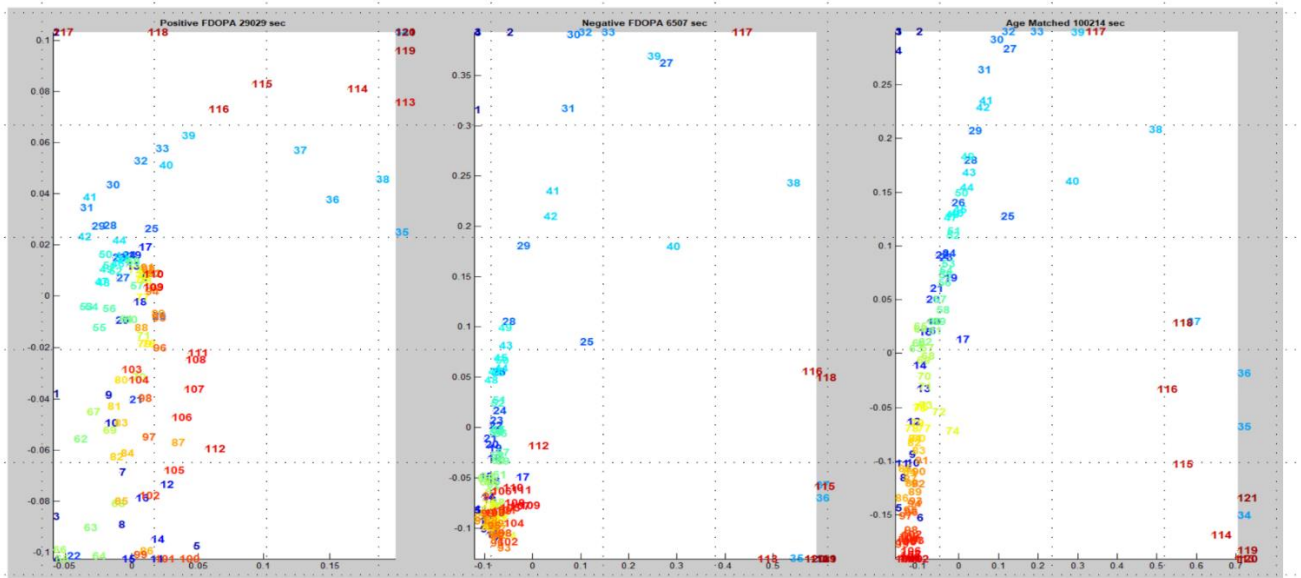

**Supplementary Figure 1.** Group connectivity representation between components extracted from a single-channel EEG, showing proximity of the components, in three groups: F-DOPA positive patients (left), F-DOPA negative patients (middle), and healthy age-matched subjects (right). The numbers represent the component number from the 121 component representation and for better visibility the numbers are also colored using the heat color map so that larger channel-numbers are red and lower channel-numbers are blue. Note that lower distance represents higher connectivity between two components. The axes of the representation have no real meaning and represent the two dimensions obtained from the optimization process (the MDS) to embed the total number of correlations onto a two-dimensional space.

## 2  F-DOPA classification Prediction

Figure 1 demonstrated the difference in connectivity between three groups of subjects. However, while the group difference may be very evident, an important question is whether this difference can be translated into individual differences which can lead to a classifier of individual subjects. To this end, we created a classifier using age-matched individuals that were not diagnosed with Parkinson's disease, and 14 F-DOPA-positive patients. We then tested the classifier on 18 subjects, of which 6 were F-DOPA negative and 12 positives. The results are depicted in Figure 3 of the main paper.

The input to the predictor is the collection of correlations between the different EEG components for each individual. These correlations are calculated during the 12-minute cognitive assessment that the patients perform. This collection of correlations is of dimension 121x120/2 which is 7260. Thus, to avoid overfitting and obtain a smaller predictor, we performed an unsupervised dimensionality reduction based on previously collected data. Specifically, using Principal Components Analysis (16), we created a standard dimensionality reduction that is used in different studies, from data of young and senior subjects all performing the same cognitive task (13,17). This corresponds to step A in the construction of the classifier (See scheme in figure 2).

Step B trains a set of logistic regression classifiers. The training data as mentioned before included 14 F-DOPA positive and 24 age-matched F-DOPA negative patients. The regressors have different ridge regularizes. We then ensemble average all the regressors (18) to obtain the final predictor.

Step C applies the obtained set of classifiers on the test data, which included 12 F-DOPA positive results and 6 F-DOPA negative results and 30 records from 20 healthy age-matched controls. First the data is projected on the previously calculated. dimensionality reduction and then passed by the full set of regressors.

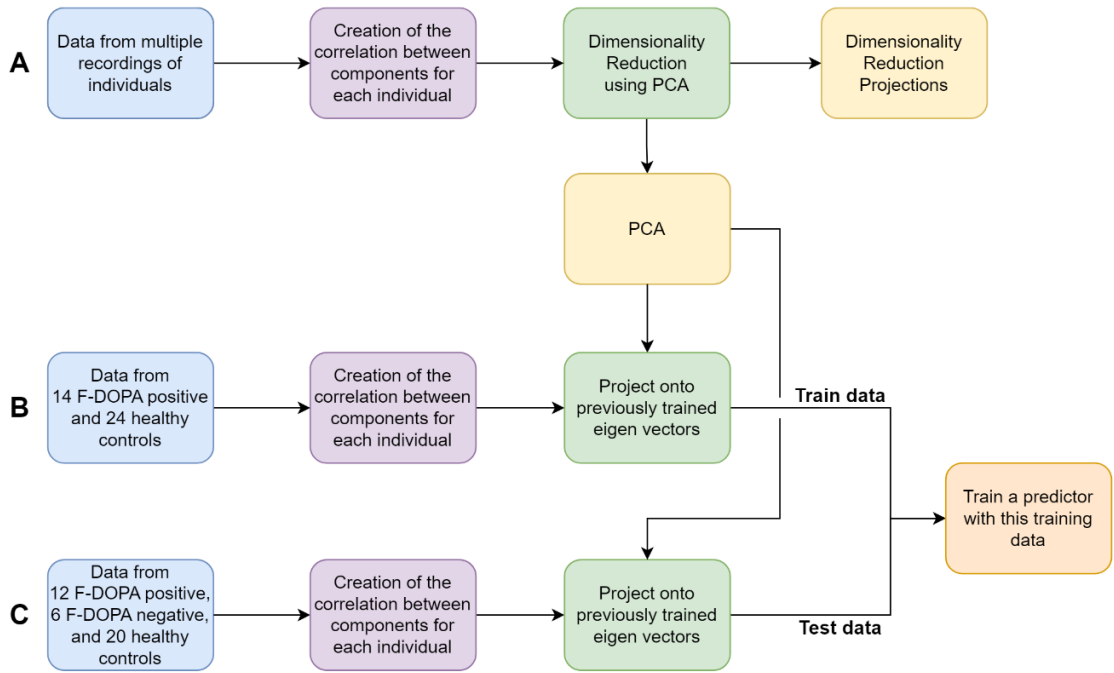

**Supplementary Figure 2.** Scheme of the classification prediction model calculations.

# *Supplementary Material B – Appendix II*

## 1   Demographic results

To ensure that the groups were well-balanced we compared some demographic characteristics of each group. Table 1 presents descriptive information regarding the motor symptoms side. The table presents the group association, gender and laterality of the motor symptoms of the patients. Data of 29 patients was collected, 3 additional patients did not have this data in their record.

| Study group | Gender | Motor symptom side | Count |
|---|---|---|---|
| Negative | Female | Bilateral | 1 |
|  |  | L | 1 |
|  |  | R | 2 |
|  | Male | Bilateral | 1 |
| Positive | Female | Bilateral | 4 |
|  |  | L | 2 |
|  |  | R | 5 |
|  | Male | Bilateral | 4 |
|  |  | L | 2 |
|  |  | R | 7 |

**Supplementary Table 1**. Motor symptom laterality information for patients holding a valid F-DOPA test result.

## 2   Bayesian results

Quantitative analysis using Bayesian Mann-Whitney U tests were performed to determine the similarity between labeled groups. This follow-up analysis was conducted using a data augmentation algorithm with 5 chains of 1000 iterations. We report the $BF_{01}$ (i.e., the null hypothesis that H0 is not different from H1) of the Bayesian U tests between controls vs. negative and positive vs. unknown, and the $BF_{10}$ (i.e., the hypothesis that H0 is different from H1) of the Bayesian U tests between control vs. positive and control vs. unknown groups.

Bayesian Mann-Whitney U tests revealed strong evidence that the predicted results of the control group differ from positive F-DOPA patients ($BF_{10} = 121.88$, W = 385, $R^2 = 1.04$), and presented moderate evidence of similarity to the negatively labeled F-DOPA group ($BF_{01} = 2.97$, W = 116, $R^2 = 1.21$). The group with unknown labels, who were all given positive predictor results, was strongly evident to differ from the control group ($BF_{10} = 149.48$, W = 550, $R^2 = 1.032$), and showed moderate evidence of similarity to the positive group ($BF_{01} = 2.145$, W = 98, $R^2 = 1$). The U tests outputs and figures are presented in tables 2-5 and figures 1-4.

|                              | BF$_{01}$ | W       | R$^2$  |
|------------------------------|-----------|---------|--------|
| Prediction normal vs. negative | 2.793     | 116.000 | 1.028  |

**Supplementary Table 2.** Bayesian Mann-Whitney U test for prediction normal vs. negative groups.

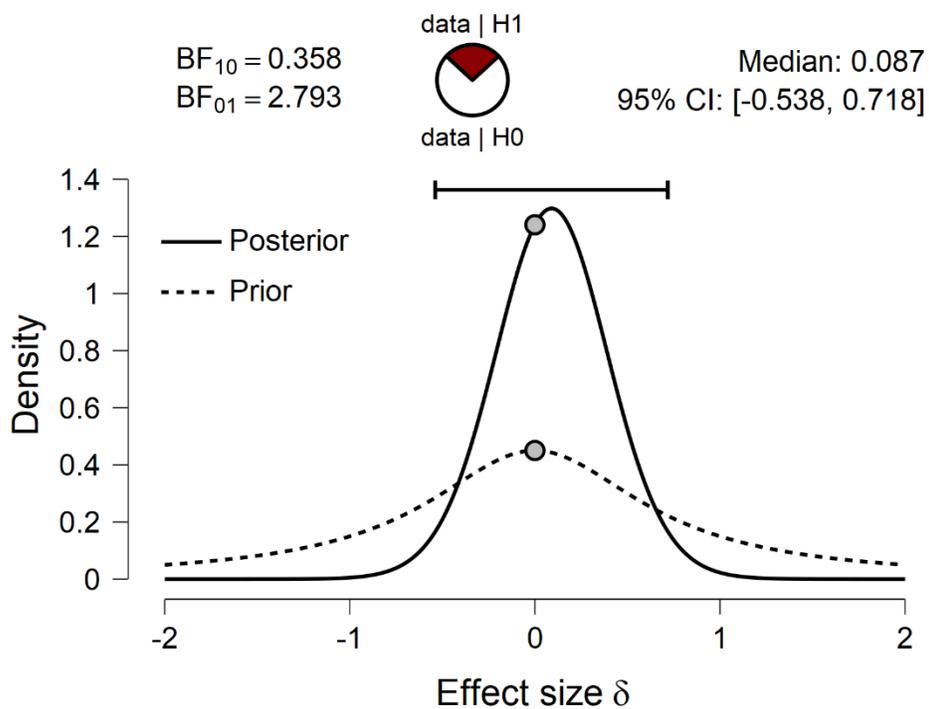

**Supplementary Figure 1.** Inferential plots prediction normal vs. negative groups prior and posterior.

|                               | $BF_{10}$ | W       | $R^2$ |
|-------------------------------|-----------|---------|-------|
| Prediction normal vs. positive | 121.881   | 385.000 | 1.043 |

**Supplementary Table 3.** Bayesian Mann-Whitney U test of prediction normal vs. positive groups.

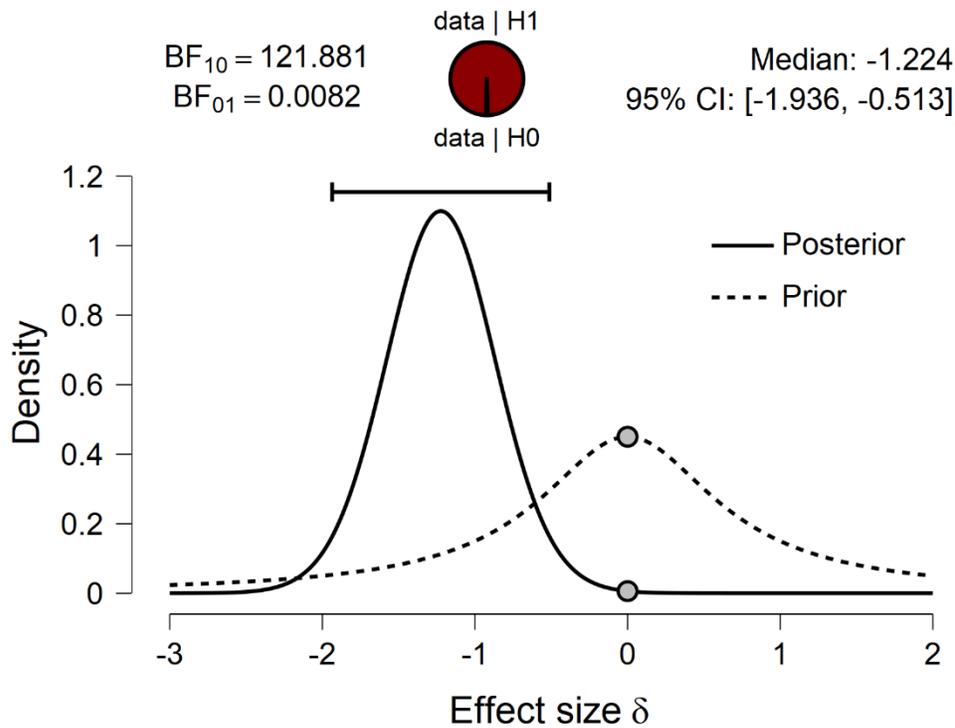

**Supplementary Figure 2.** Inferential plots prediction normal vs. positive groups prior and posterior.

|  | BF$_{10}$ | W | R$^2$ |
|---|---|---|---|
| Prediction normal vs. unknown | 149.481 | 550.000 | 1.032 |

**Supplementary Table 4.** Bayesian Mann-Whitney U test for prediction normal vs. unknown groups.

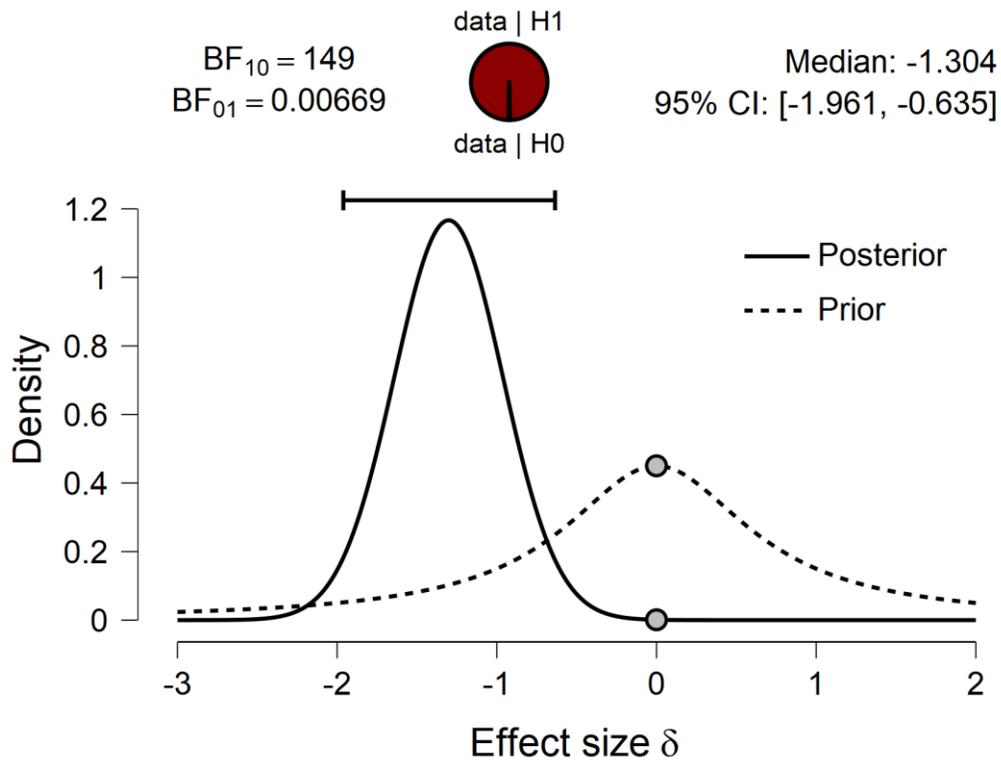

**Supplementary Figure 3.** Inferential plots prediction normal vs. unknown groups prior and posterior.

|                                  | $BF_{01}$ | W      | $R^2$  |
|----------------------------------|-----------|--------|--------|
| Prediction positive vs. unknown  | 2.145     | 98.000 | 1.000  |

**Supplementary Table 5.** Bayesian Mann-Whitney U test for prediction positive vs. unknown groups.

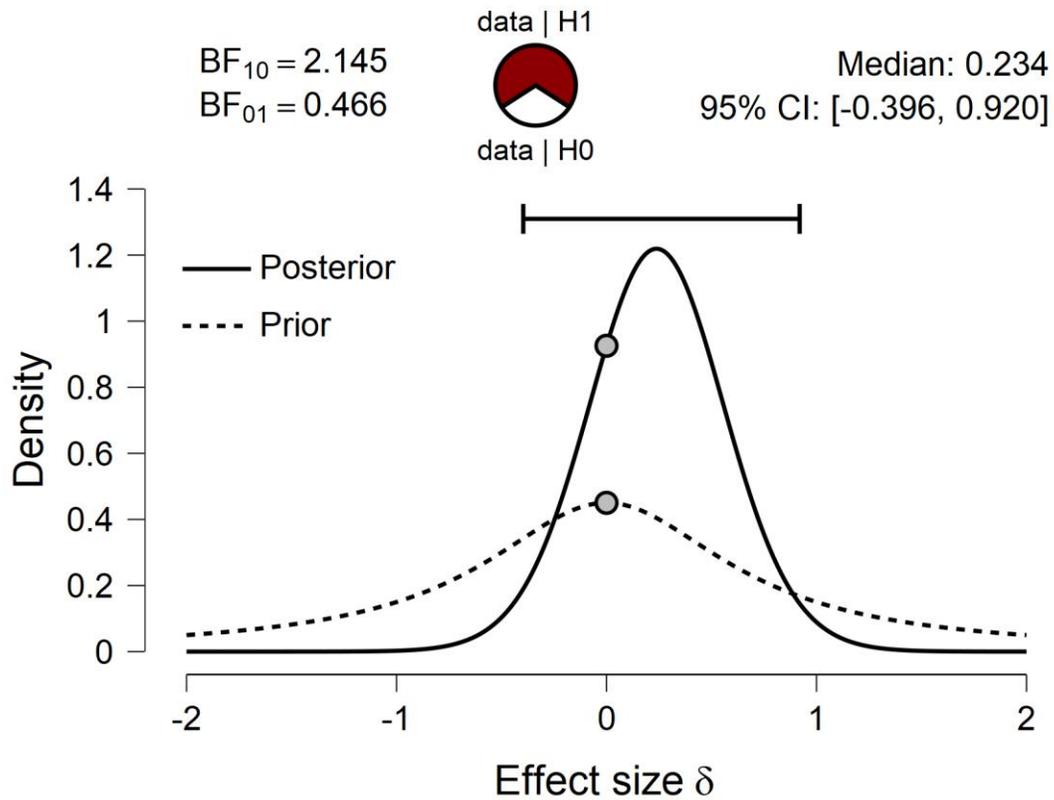

**Supplementary Figure 4.** Inferential plots prediction positive vs. unknown groups prior and posterior.